\documentclass[aps,manuscript,showpacs,showkeys,superscriptaddress]{revtex4-1}
\usepackage{graphicx}% Include figure files
\usepackage{epsfig}  % Include figure files
\usepackage{epsf}    % Include figure files
\usepackage{dcolumn}% Align table columns on decimal point
\usepackage{bm}% bold math
\usepackage{dcolumn}% Align table columns on decimal point
\usepackage{textcomp}% Align table columns on decimal point
\usepackage[tbtags]{amsmath}
\usepackage{amsfonts}
\usepackage{float}
\usepackage{subfig}
\usepackage[]{hyperref}
  \hypersetup{
  unicode=false,          % non-Latin characters in Acrobat?s bookmarks
  pdftoolbar=true,        % show Acrobat?s toolbar?
  pdfmenubar=true,        % show Acrobat?s menu?
  pdffitwindow=true,     % window fit to page when opened
  pdfstartview={FitH},    % fits the width of the page to the window
  pdfsubject={Getting the best out of T2K and NOvA},   % subject of the document
  pdfnewwindow=true,      % links in new window
  pdfcreator={RevTeX},
  colorlinks=true,       % false: boxed links; true: colored links
  linkcolor=red,          % color of internal links
  citecolor=blue,        % color of links to bibliography
  urlcolor=blue,           % color of external links
  }
\usepackage{hypcap}

\newcommand{\beq}{\begin{equation}}
\newcommand{\eeq}{\end{equation}}
\newcommand{\beqa}{\begin{eqnarray}}
\newcommand{\eeqa}{\end{eqnarray}}

\newcommand{\tx}{{\theta_{12}}}
\newcommand{\ty}{{\theta_{13}}}
\newcommand{\tz}{{\theta_{23}}}

\newcommand{\dhat}{{\hat{\Delta}}}
\newcommand{\ahat}{{\hat{A}}}

\newcommand{\dl}{{\Delta_{31}}}
\newcommand{\ds}{{\Delta_{21}}}

\newcommand{\dcp}{{\delta_{CP}}}
\newcommand{\pmue}{P_{\mu e}}

\newcommand{\pmuebar}{P_{\bar{\mu}\bar{e}}}
\newcommand{\dchsq}{\Delta \chi^2}
\newcommand{\nova}{NO$\nu$A~}

\newcommand{\be}{\begin{equation}}
\newcommand{\ee}{\end{equation}}
\newcommand{\ba}{\begin{eqnarray}}
\newcommand{\ea}{\end{eqnarray}}

\begin{document} 
\title{Looking for Lorentz invariance violation (LIV) in the latest long baseline accelerator neutrino oscillation data}

%% %simple case: 2 authors, same institution
%% \author{A. Uthor}
%% \author{and A. Nother Author}
%% \affiliation{Institution,\\Address, Country}

% more complex case: 4 authors, 3 institutions, 2 footnotes
\author{Ushak Rahaman}
%\author[c]{S. Econd,}
%\author[a,2]{T. Hird\note{Also at Some University.}}
%\author[a,2]{and Fourth}

% The "\note" macro will give a warning: "Ignoring empty anchor..."
% you can safely ignore it.

\affiliation{Centre for Astro-Particle Physics (CAPP) and Department of Physics, University of Johannesburg, PO Box 524, Auckland Park 2006, South Africa}

% e-mail addresses: one for each author, in the same order as the authors
%\emailAdd{ushakr@uj.ac.za}

\date{Received: date / Revised version: date}

\begin{abstract}
In this paper, we have analysed the latest data from \nova and T2K with the Lorentz invariance violation along with the standard oscillation hypothesis. We have found that the \nova data cannot distinguish between the two hypotheses at $1\, \sigma$ confidence level. T2K data and the combined data analysis excludes standard oscillation at $1\, \sigma$. All three cases do not have any hierarchy sensitivity when analysed with LIV. There is a mild tension between the two experiments, when analysed with LIV, as $\tz$ at \nova best-fit is at higher octant but the same for T2K is at lower octant. The present data from accelerator neutrino long baseline experiments lose octant determination sensitivity when analysed with LIV. The tension between the two experiments is also reduced when the data are analysed with LIV.
\end{abstract}

\maketitle
\flushbottom

\section{Introduction}
\label{sec:intro}
The three flavour neutrino oscillation phenomenon is parameterised by the unitary Pontecorvo-Maki-Nakagawa-Sakata (PMNS) mixing matrix:
\begin{equation}
U=\left[
\begin{array}{ccc}
c_{13}c_{12} & s_{12}c_{13} & s_{13}e^{-i\dcp}\\
-s_{12}c_{23}-c_{12}s_{23}s_{13}e^{i\dcp} & c_{12}c_{23}-s_{12}s_{23}s_{13}e^{i\dcp} & s_{23}c_{13}\\
s_{12}s_{23}-s_{13}c_{12}c_{23}e^{i\dcp} & -c_{12}s_{23}-s_{13}c_{23}s_{12}e^{i\dcp} & c_{23}c_{13}\\
\end{array}
\right],
\end{equation}
where $c_{ij}=\cos \theta_{ij}$ and $s_{ij}=\sin \theta_{ij}$. The neutrino oscillation probabilities depend on three mixing angles: $\tx$, $\ty$, and $\tz$; two independent mass squared differences: $\ds=m_{2}^{2}-m_{1}^{2}$ and $\dl=m_{3}^{2}-m_{1}^{2}$, where $m_1$, $m_2$ and $m_3$ are the masses of the neutrino mass eigenstates
$\nu_1$, $\nu_2$ and $\nu_3$ respectively; and a CP violating phase: $\dcp$. Among these parameters, $\tx$ and $\ds$ have been measured in solar neutrino experiments \cite{Bahcall:2004ut, Ahmad:2002jz}. $\sin^2 2\tz$ and $|\dl|$ has been obtained by measuring the $\nu_\mu$ survival probabilities in accelerator neutrino long baseline experiment MINOS \cite{Kyoto2012MINOS}.Recently reactor neutrino experiments have measured non-zero value of $\ty$ \cite{An:2012eh, Ahn:2012nd, Abe:2011fz}. In table \ref{bestfit}, we have noted down the current best-fit values along with the $1\, \sigma$ uncertainties of the oscillation parameters. 
\begin{table}
  \begin{center}
\begin{tabular}{|c|c|c|}
%  \showrowcolors
  \hline
  Parameters & NH \phantom{foo}&
  IH\\
  \hline %\hiderowcolors
  $\tx/^o$ & $33.44^{+0.77}_{-0.74}$ & $33.45^{+0.78}_{-0.75}$\\
  \hline
  $\tz/^o$ & $49.2^{+0.9}_{-1.2}$ & $49.3^{+0.9}_{-1.1}$\\
  \hline
  $\ty/^o$ & $8.57^{+0.12}_{-0.12}$ & $8.60^{+0.12}_{-0.12}$\\
  \hline
  $\dcp/^o$  & $197^{+24}_{-24}$ & $282^{+26}_{-30}$ \\
  \hline
 $ \frac{\ds}{10^{-5}\, {\rm eV}^2}$ & $7.42^{+0.21}_{-0.20}$ & $7.42^{+0.21}_{-0.20}$ \\
 \hline
  $ \frac{\Delta_{3l}}{10^{-3}\, {\rm eV}^2}$ & $2.517^{+0.026}_{-0.028}$ & $-2.498^{+0.028}_{-0.028}$ \\
 \hline
\end{tabular}
\end{center}
 \caption{Global best-fit values of neutrino oscillation parameters \cite{nufit, Esteban:2020cvm}. $\Delta_{3l}=\dl>0$ ($\Delta_{32}<0$) for NH (IH).}
  \label{bestfit}
\end{table}

The current unknowns are the sign of $\dl$, octant of $\tz$ and the CP violating phase $\dcp$.There can be two possible mass orderings
for the masses of the neutrino mass eigenstates, depending on the sign of $\dl$: normal hierarchy (NH), which implies
$m_3>>m_2>m_1$, and inverted hierarchy (IH), which implies $m_2>m_1>>m_3$ \cite{deSalas:2018bym}. It is expected that the current accelerator based long baseline neutrino oscillation experiments \nova \cite{Ayres:2004js} and T2K \cite{Itow:2001ee} will measure these unknowns by measuring the $\nu_\mu \to \nu_e$ oscillation probabilities in presence of matter effect. Recently both the experiments have published their latest analysis. The best-fit values for \nova is $\sin^2 \tz =
0.57^{+0.04}_{-0.03}$ and $\dcp=0.82\pi$ for NH \cite{Himmel:2020}. For T2K, the best-fit values are $\sin^2
\tz=0.53^{+0.03}_{-0.04}$ for both mass hierarchies, 
and $\dcp/\pi=-1.89^{+0.70}_{-0.58}$ ($-1.38^{+0.48}_{-0.54}$) for
normal (inverted) hierarchy \cite{Dunne:2020}. Therefore, there is a moderate tension between the outcomes of the two experiments. The measured best-fit $\dcp$ values of both the experiments are far apart. Moreover, there is no overlap between the allowed regions on $\sin^2 \tz-\dcp$ plane at $1\, \sigma$ confidence level (C.L.). Although the individual experiments prefer NH over IH, their combined analysis has the best-fit point at IH \cite{Kelly:2020fkv}.

Apart from the unknown standard oscillation parameters, these experiments will also investigate about the possibility of beyond standard model (BSM) physics.A large number of studies have been done about exploring BSM physics with long baseline neutrino oscillation experiments \cite{Arguelles:2019xgp}. Recently non-unitary neutrino mixing \cite{Miranda:2019ynh} and non-standard neutrino interaction during propagation through matter \cite{Chatterjee:2020kkm, Denton:2020uda} have been used to resolve the tension between \nova and T2K.
It is important to test other BSM physics models to resolve the tension as well. 

Neutrino oscillation requires neutrinos to be massive albeit extremely light. This curious and interesting characteristic makes neutrino oscillation the first experimental signature of BSM physics. Without loss of any generality, SM can be considered as the low energy effective theory derived from a more general theory governed by Planck mass ($M_P\simeq 10^{19}\, {\rm GeV}$). This more fundamental theory unifies gravitational interactions along with strong, weak and electro-magnetic interactions. There exists theoretical models \cite{Kostelecky:1988zi, Kostelecky:1989jp, Kostelecky:1991ak, Kostelecky:1994rn, Kostelecky:1995qk} which include spontaneous Lorentz invariance violation (LIV) and CPT violations in that more complete framework at Planck scale. At the observable low energy, these violations can give rise to minimal extension of SM through perturbative terms suppressed by $M_P$. Particles and anti-particles have same mass and lifetime due to CPT invariance. Any observed difference between masses or lifetimes of particles and anti-particles would be a signal for CPT violation. The present upper limit on CPT violation from kaon system is $|m_{K^0}-m_{\bar{K}_0}|/m_K<6 \times 10^{-18}$ \cite{Tanabashi:2018oca}. Since, kaons are Bosons and the natural mass term appearing in the Lagrangian is mass squared term, the above constraints can be rewritten as $|m_{K^0}^{2}-m_{\bar{K}_0}^{2}|<0.25\, {\rm eV}^2$. Current neutrino oscillation data provide the bounds $|\ds-\bar{\Delta}_{21}|<5.9 \times 10^{-5}\, {\rm eV}^2$ and $|\dl-\bar{\Delta}_{31}|<1.1 \times 10^{-3}\, {\rm eV}^2$ \cite{Ohlsson:2014cha}. If these differences are non-zero and they are manifestation of some kind of CPT violating effects, they can induce changes in neutrino oscillation probabilities \cite{Kostelecky:2003cr, Diaz:2011ia, Kostelecky:2004hg, Katori:2006mz}. Various studies have been done about the LIV/CPT violation with neutrinos \cite{Dighe:2008bu, Barenboim:2009ts, Rebel:2013vc, deGouvea:2017yvn, Barenboim:2017ewj, Barenboim:2018ctx, Majhi:2019tfi, Giunti:2010zs, Datta:2003dg, Chatterjee:2014oda, Koranga:2014dua, Diaz:2016fqd, Hooper:2005jp, Tomar:2015fha, Liao:2017yuy, Agarwalla:2019rgv}. Several neutrino oscillation experiments have looked for LIV/CPT violations and put on constraints on the LIV/CPT violating parameters \cite{Auerbach:2005tq, Adamson:2008aa, Adamson:2012hp, Aguilar-Arevalo:2018gpe, Abe:2012gw, Abe:2014wla, Aartsen:2017ibm, Abe:2017eot}. Ref.~\cite{Kostelecky:2008ts} includes the list of constraints on all the relevant LIV/CPT violating parameters. But till date no study has been made to look for LIV/CPT violation in the long baseline accelerator neutrino oscillation experiments {\bf data}. In this paper, we will consider the minimal extension of SM that violates Lorentz invariance as well as CPT symmetry. We will test the model with the latest data from \nova and T2K and try to see if there is any hint of CPT violating LIV in the individual as well as combined data set and whether the tension between the two experiments can be resolved with the help of this new physics hypothesis. 

In section \ref{theory}, we will discuss the theoretical framework of LIV in neutrino oscillation and present the comparison between oscillation probabilities with and without LIV. In section \ref{analysis}, we will discuss the details method of our analysis and in section \ref{result} we will present our results after analysing data from \nova and T2K. The conclusion will be drawn in section \ref{conclusion}.
\section{Theoretical framework}
\label{theory}
The Lorentz invariance violating neutrinos and anti-neutrinos can be described by the effective Lagrangian \cite{Kostelecky:2003cr, Kostelecky:2011gq}
\begin{equation}
    \mathcal{L}=\bar{\Psi}_A\left(i \gamma_\mu \partial_\mu \delta_{AB}-M_{AB}+\hat{\mathcal{Q}}_{AB}\right)\Psi_B+{\rm h.c.}.
    \label{LIV-lag}
\end{equation}
$\Psi_{A(B)}$ is a $2N$ dimensional spinor containing $\psi_{\alpha(\beta)}$, which is a spinor field with $\alpha(\beta)$ ranging over $N$ spinor flavours, and their charge conjugates given by $\psi_{\alpha(\beta)}^{C}=C\bar{\psi}_{\alpha(\beta)}^T$. Therefore, $\Psi_{A(B)}$ can be expressed as
\begin{equation}
    \Psi_{A(B)}=\left(\psi_{\alpha(\beta)}, \psi_{\alpha(\beta)}^{C}\right)^T.
\end{equation}
$\hat{\mathcal{Q}}$ in eq.~\ref{LIV-lag} is a generic Lorentz invariance violating operator. The first term in the right side of eq.~\ref{LIV-lag} is the kinetic term, the second term is the mass term involving the mass matrix $M$ and the third term gives rise to the LIV effect. $\hat{\mathcal{Q}}$ is small and perturbative in nature.

We will restrict ourselves only to the renormalizable Dirac couplings in the theory, i.e. terms only with mass dimension $\leq 4$ will be incorporated. Doing so, one can write the Lorentz invariance violating Lagrangian in the flavour basis as \cite{Kostelecky:2003cr}
\begin{equation}
    \mathcal{L}_{\rm LIV}= -\frac{1}{2}\left[a^{\mu}_{\alpha \beta}\bar{\psi}_\alpha \gamma_\mu \psi_\beta+b^{\mu}_{\alpha \beta}\bar{\psi}_\alpha \gamma_5 \gamma_\mu \psi_\beta-i c_{\alpha \beta}^{\mu \nu}\bar{\psi}_\alpha \gamma_\mu \partial_\nu \psi_\beta-i d_{\alpha \beta}^{\mu \nu}\bar{\psi}_\alpha \gamma_\mu \gamma_5 \partial_\nu \psi_\beta\right],
\end{equation}
where $a^{\mu}_{\alpha \beta}$, $b^{\mu}_{\alpha \beta}$, $c^{\mu \nu}_{\alpha \beta}$ and $d^{\mu \nu}_{\alpha \beta}$ Lorentz invariance violating parameters. Since, only left handed neutrinos are present in SM, the observable effects in the neutrino oscillation experiments can be parameterized as
\begin{equation}
    \left(a_L\right)^{\mu}_{\alpha \beta}=\left(a+b\right)^{\mu}_{\alpha \beta}, \left(c_L\right)^{\mu \nu}_{\alpha \beta}=\left(c+d\right)^{\mu \nu}_{\alpha \beta}.
\end{equation}
These are constant Hermitian matrices which can modify the standard Hamiltonian in vacuum. The first combination involves CPT violation, where as the second combination is the CPT conserving Lorentz invariance violating neutrinos. In this paper, we will consider only direction independent isotropic terms, and hence we will only consider the $\mu=\nu=0$. From now on, for simplicity, we will call $a_{\alpha \beta}^{0}$ terms as $a_{\alpha \beta}$ and $c_{\alpha \beta}^{00}$ term as $c_{\alpha \beta}$. Taking into account only these isotropic LIV terms, the neutrino Hamiltonian with LIV effect becomes:
\begin{equation}
    H=H_{\rm vac}+H_{\rm mat}+H_{\rm LIV},
    \label{Ham}
\end{equation}
where 
\begin{equation}
    H_{\rm vac}=\frac{1}{2E}U\left[
\begin{array}{ccc}
m_{1}^{2} & 0 & 0\\
0 & m_{2}^{2} & 0\\
0 & 0 & m_{3}^{2}\\
\end{array}
\right]U^\dagger; H_{\rm mat}=\sqrt{2}G_FN_e \left[
\begin{array}{ccc}
1 & 0 & 0\\
0 & 0 & 0\\
0 & 0 & 0\\
\end{array}
\right];
\end{equation}
\begin{equation}
    H_{\rm LIV}=\left[
\begin{array}{ccc}
a_{ee} & a_{e\mu} & a_{e\tau}\\
a_{e\mu}^{*} & a_{\mu \mu} & a_{\mu \tau}\\
a_{e\tau}^{*} & a_{\mu\tau}^{*} & a_{\tau \tau}\\
\end{array}
\right] -\frac{4}{3}E\left[
\begin{array}{ccc}
c_{ee} & c_{e\mu} & c_{e\tau}\\
c_{e\mu}^{*} & c_{\mu \mu} & c_{\mu \tau}\\
c_{e\tau}^{*} & c_{\mu\tau}^{*} & c_{\tau \tau}\\
\end{array}
\right].
\end{equation}
$G_F$ is the Fermi coupling constant and $N_e$ is the electron density along the neutrino path. The $-4/3$ in front of the second term arises due to non observability of the Minkowski trace of the CPT conserving LIV term $c_L$ which relates $xx$, $yy$, and $zz$ component to the $00$ component \cite{Kostelecky:2003cr}. The effects of $a_{\alpha \beta}$ are proportional to the baseline $L$ and those of $c_{\alpha \beta}$ are proportional to $LE$. In this paper, we will consider the effects of CPT violating LIV parameters $a_{\alpha \beta}$ only. 

It is noteworthy that the Hamiltonian due to LIV is analogous to that with neutral current (NC) non standard interaction (NSI) during the propagation of neutrinos through matter
\begin{equation}
    H^\prime=H_{\rm vac}+H_{\rm mat}+H_{\rm NSI},
\end{equation}
where
\begin{equation}
    H_{\rm NSI}=\sqrt{2}G_FN_e\left[
\begin{array}{ccc}
\epsilon_{ee} & \epsilon_{e\mu} & \epsilon_{e\tau}\\
\epsilon_{e\mu}^{*} & \epsilon_{\mu \mu} & \epsilon_{\mu \tau}\\
\epsilon_{e\tau}^{*} & \epsilon_{\mu\tau}^{*} & \epsilon_{\tau \tau}\\
\end{array}
\right].
\end{equation}
$\epsilon_{\alpha \beta}$ are the strength of NSI. Thus, a relation between CPT violating LIV and NSI can be found by following equation \cite{Diaz:2015dxa}:
\begin{equation}
    \epsilon_{\alpha \beta}=\frac{a_{\alpha \beta}}{\sqrt{2}G_F N_e}.
    \label{eps-a}
\end{equation}

In this work, we will consider the effects of parameters $a_{e \mu}=|a_{e \mu}|e^{i \phi_{e\mu}}$ and $a_{e \tau}=|a_{e \tau}|e^{i \phi_{e\tau}}$, because these two parameters have the highest influences on $\nu_\mu \to \nu_e$ oscillation probability \cite{Agarwalla:2019rgv}, which is responsible for octant, $\dcp$, and hierarchy sensitivity of long baseline accelerator neutrino like \nova and T2K. Since, we are mostly concerned about determining these unknown standard oscillation parameters in the long baseline accelerator neutrino experiment, we fixed all other LIV parameters, except $a_{e\mu}$ and $a_{e\tau}$, to zero. It implies that in this paper, the contribution from Lorez invariance violation in the \nova and T2K experiments is coming only from the CPT violating Lorentz violation and mostly in the appearance channels. The $\nu_\mu$ and $\bar{\nu}_\mu$ disappearance channels still conserve Lorentz invariance. The current constraint on these parameters from Super-kamiokande experiment at $95\%$ confidence level (C.L.) is \cite{Abe:2014wla} 
\begin{equation}
    |a_{e\mu}|<2.5\times 10^{-23}\, {\rm GeV};\, |a_{e\tau}|<5\times 10^{-23}\, {\rm GeV}
\end{equation}

The $\nu_\mu \to \nu_e$ oscillation probability in presence of LIV parameters $a_{e\mu}$ and $a_{e\tau}$ can be written in the similar way as the oscillation probability in presence of NSI parameters
$\epsilon_{e\mu}$ and $\epsilon_{e\tau}$ \cite{Kikuchi:2008vq, Agarwalla:2016fkh, Masud:2018pig}:
\begin{equation}
    P_{\mu e}^{\rm SM+LIV}\simeq P_{\mu e} (\rm SM)+P_{\mu e} (a_{e\mu})+P_{\mu e}(a_{e\tau}).
    \label{pmue}
\end{equation}
The first term in eq.~\ref{pmue} is the oscillation probability in the presence of standard matter effect. It can be written as \cite{Cervera:2000kp}
 \begin{eqnarray}
  &\pmue (\rm SM)& = \sin^2 2 \ty \sin^2 \tz\frac{\sin^2\dhat(1-\ahat)}{(1-\ahat)^2}\nonumber\\
  &+&\alpha \cos \ty \sin2\tx \sin 2\ty \sin 2\tz \cos(\dhat+\dcp)\nonumber\\
 &&\frac{\sin\dhat \ahat}{\ahat}
  \frac{\sin \dhat(1-\ahat)}{1-\ahat}\nonumber\\
  &+& \alpha^2 \sin^2 2\theta_{12}\cos^2 \theta_{13}\cos^2\theta_{23}\frac{\sin^2 \dhat \ahat}{\ahat^2},
  \label{pme1}
   \end{eqnarray}
   where $\alpha=\frac{\ds}{\dl}$, $\dhat=\frac{\dl L}{4E}$
and $\ahat=\frac{A}{\dl}$. $A$ is the Wolfenstein matter term \cite{msw1}, given by 
$A=2\sqrt{2}G_FN_eE$, where $E$ is the neutrino beam energy and $L$ is the length of the baseline. 

For the second and third terms in eq.~\ref{pmue}, describing the effects of $a_{e\mu}$ and $a_{e\tau}$ respectively, we follow the similar approach of NSI, described in references \cite{Kikuchi:2008vq, Agarwalla:2016fkh, Masud:2018pig}, and replace $\epsilon_{\alpha \beta}$ terms by $a_{\alpha \beta}$ terms according to eq.~\ref{eps-a}. Doing so, we can write
\begin{eqnarray}
    P_{\mu e}(a_{e\beta})&=&\frac{4|a_{e\beta}|\ahat \dhat\sin \theta_{13}\sin 2\theta_{23}\sin \dhat}{\sqrt{2}G_F N_e}\left[Z_{e\beta}\sin(\dcp+\phi_{e\beta})+W_{e\beta}\cos(\dcp+\phi_{e\beta}) \right] \nonumber \\
    &=& \frac{4|a_{e\beta}|L \sin \theta_{13}\sin 2\theta_{23}\sin \dhat}{2}\left[Z_{e\beta}\sin(\dcp+\phi_{e\beta})+W_{e\beta}\cos(\dcp+\phi_{e\beta}) \right]\nonumber\\
    \label{pmue-a}
\end{eqnarray}
where $\beta=\mu,\, \tau$; 
\begin{eqnarray}
 Z_{e\beta}&=&-\cos \theta_{23} \sin \dhat,\, {\rm if}\, \beta=\mu \nonumber\\
 &=& \sin \theta_{23}\sin \dhat,\, {\rm if}\, \beta=\tau
\end{eqnarray}
and
\begin{eqnarray}
 W_{e\beta}&=&\cos \theta_{23} \left(\frac{\sin^2\tz \sin \dhat}{\cos^2 \tx \dhat}+\cos \dhat\right),\, {\rm if}\, \beta=\mu \nonumber\\
 &=& \sin \theta_{23} \left(\frac{\sin\dhat}{\dhat}-\cos \dhat\right),\, {\rm if}\, \beta=\tau.
\end{eqnarray}
From, eq.~\ref{pmue-a}, it can be seen that the LIV effects considered in this paper are matter independent.

The oscillation probability $\pmuebar$ for anti-neutrino can be calculated from equations~\ref{pme1} and \ref{pmue-a} by substituting $A\to -A$, $\dcp \to -\dcp$, $|a_{e\beta}|\to -|a_{e\beta}|$ and $\phi_{e\beta}\to -\phi_{e\beta}$, where $\beta=\mu,\, \tau$.

In our analysis, however, we have used GLoBES \cite{Huber:2004ka, Huber:2007ji} to calculate the oscillation probability. To do so, we modified the probability code of the software to include LIV. After that, GLoBES is capable of calculating the oscillation probability without the approximations required to derive equations \ref{pmue}-\ref{pmue-a}. The evolution equation for a neutrino state $|\nu>=\left(|\nu_e>,|\nu_\mu>,|\nu_\tau>\right)^T$ travelling a small distance $x$ can be written in presence of LIV as
\begin{equation}
    i \frac{d}{dx}|\nu>=H|\nu>.
\end{equation}
$H$ is the Hamiltonian from eq.~\ref{Ham}. The oscillation probability of $\nu_\mu \to \nu_e$ after travelling through a distance $L$ can be written as 
\begin{equation}
    \pmue^{\rm SM+LIV}=|<\nu_e|e^{-iHL}|\nu_\mu>|^2.
\end{equation}

\begin{figure}[htbp]
\centering
\includegraphics[width=1.0\textwidth]{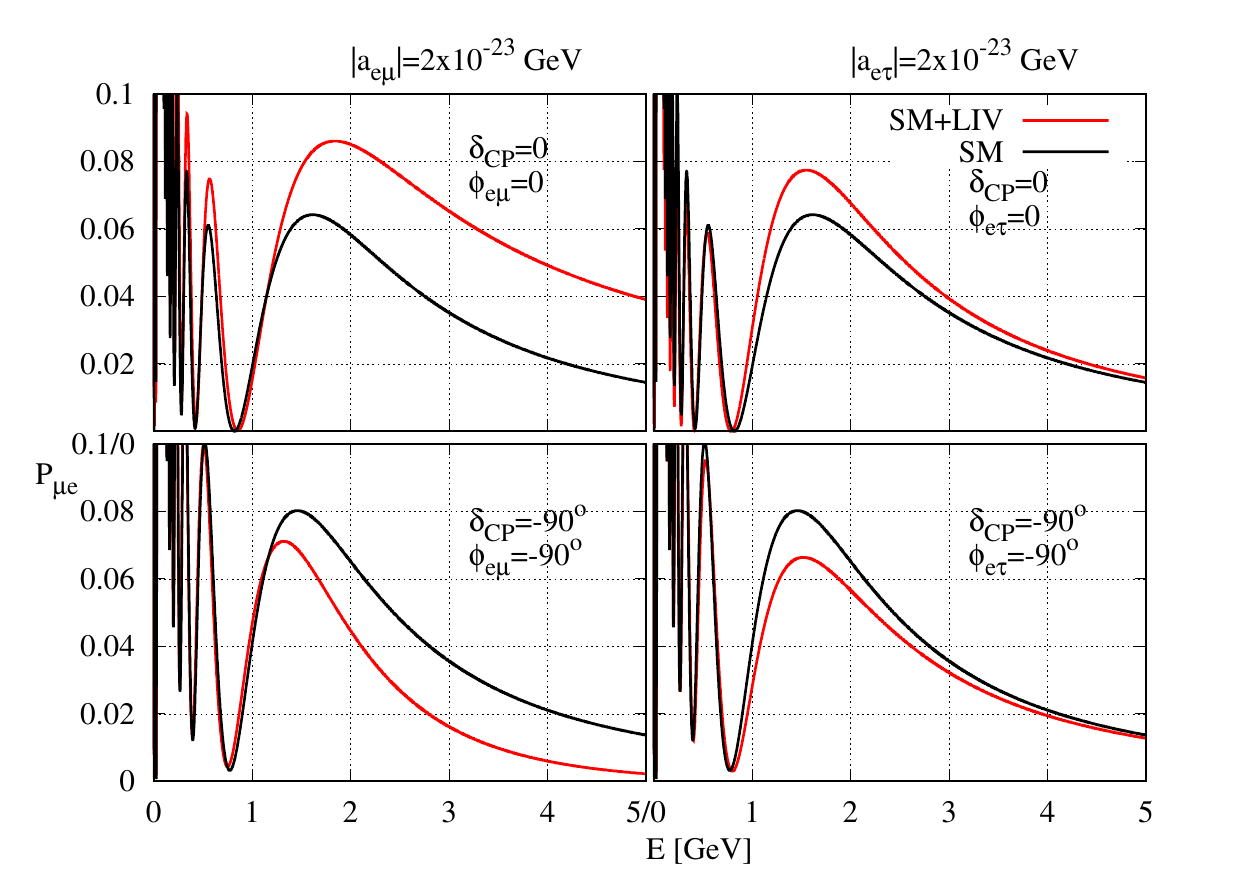}
\caption{\footnotesize{Effect of $|a_{e\mu}|(|a_{e\tau}|)=2\times 10^{-23}$ GeV on the oscillation probability for NH of \nova in the left (right panel). The black line shows the oscillation probability with standard matter effect. The standard oscillation parameter values have been fixed to their best-fit values \cite{nufit, Esteban:2020cvm}. The red line indicates the oscillation probability with LIV effect. Values of $\dcp$, $\phi_{e\mu}$ and $\phi_{e\tau}$ have been mentioned in the panels.}}
\label{prob-nova}
\end{figure}

In fig.~\ref{prob-nova}, we have shown the effects of $|a_{e\mu}|$ and $|a_{e\tau}|$ on the oscillation probability for NH of \nova for different values of $\dcp$ and $\phi_{e\mu}$, $\phi_{e\tau}$. To generate these plots, we have fixed the standard oscillation parameter values to their best-fit values taken from \cite{nufit, Esteban:2020cvm}. It is obvious that there is a large difference between standard oscillation and oscillation with LIV at probability level.This difference makes our motivation to test LIV with oscillation data from long baseline experiments even stronger. It can be observed that $\phi_{e \mu}$ $(\phi_{e\tau})$ have quite opposite effects on $P_{\mu e}$ compared to that of $\dcp$. 

\begin{figure}[htbp]
\centering
\includegraphics[width=1.0\textwidth]{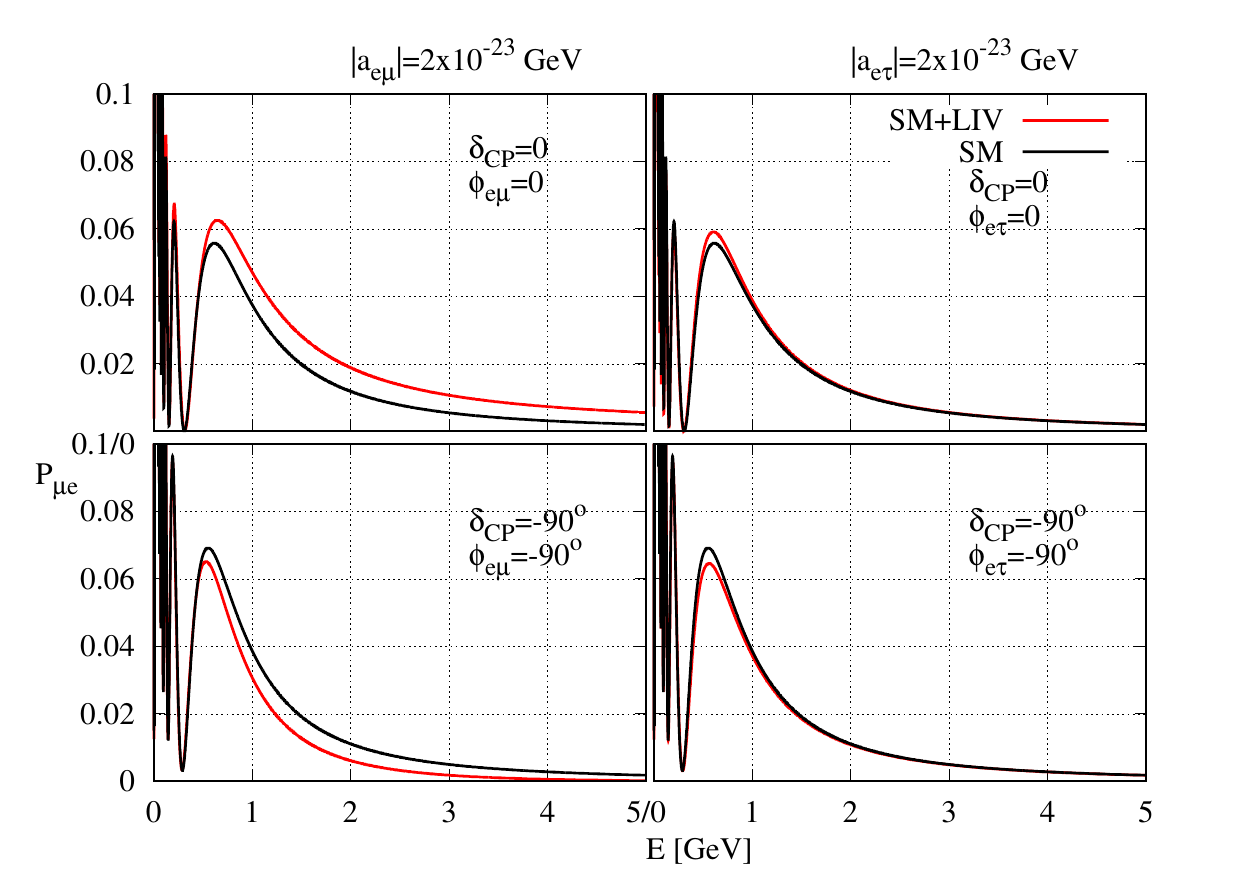}
\caption{\footnotesize{Effect of $|a_{e\mu}|(|a_{e\tau}|)=2\times 10^{-23}$ GeV on the oscillation probability for NH of T2K in the left (right panel). The black line shows the oscillation probability with standard matter effect. The standard oscillation parameter values have been fixed to their best-fit values \cite{nufit, Esteban:2020cvm}. The red line indicates the oscillation probability with LIV effect. Values of $\dcp$, $\phi_{e\mu}$ and $\phi_{e\tau}$ have been mentioned in the panels.}}
\label{prob-t2k}
\end{figure}

In fig.~\ref{prob-t2k}, we have shown the similar comparison for T2K. It is obvious that the effect of LIV is less prominent in case of T2K than it was for NO$\nu$A. It is expected because the CPT violating LIV effect is proportional to the length of the baseline and T2K baseline (295 km) is very small compared to that of \nova (810 km).

\begin{figure}[htbp]
\centering
\includegraphics[width=1.0\textwidth]{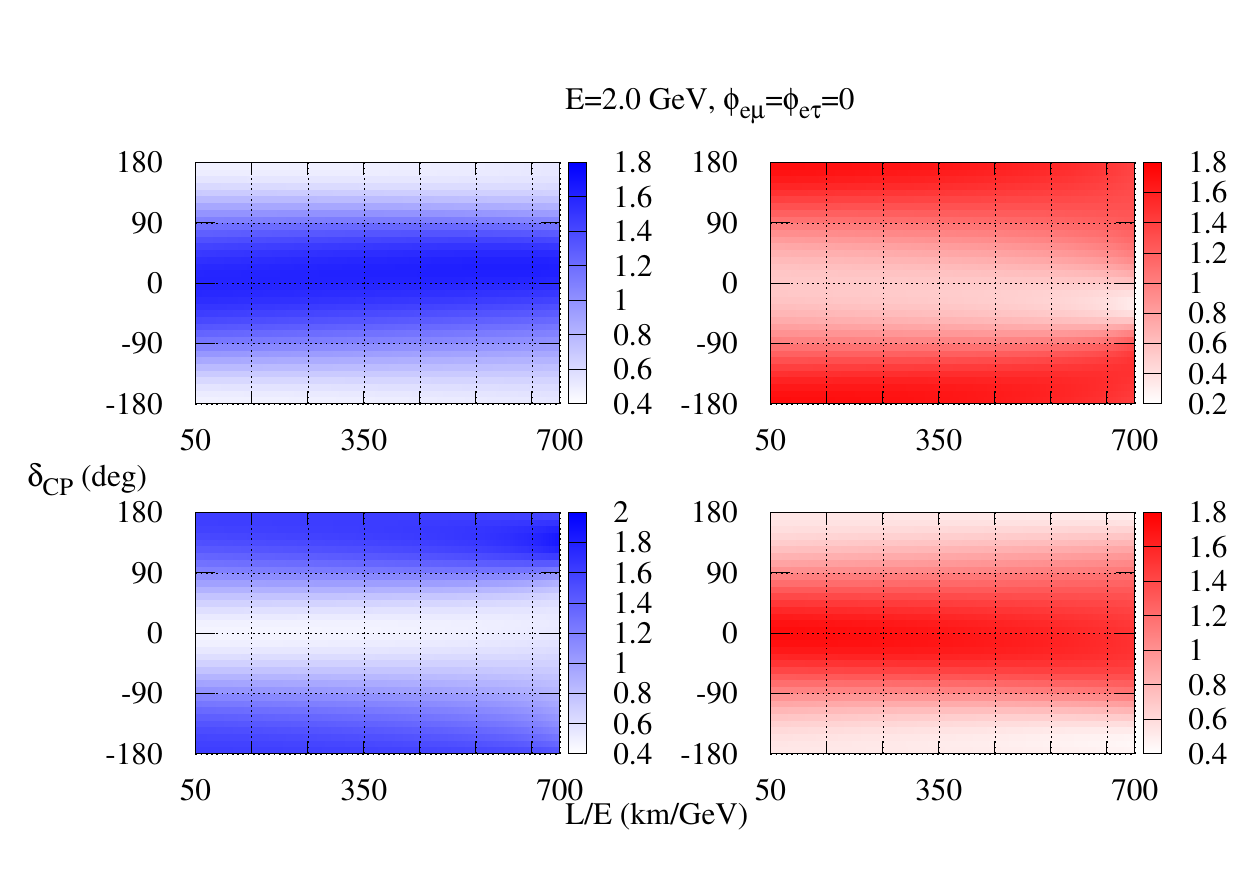}
\caption{\footnotesize{Ratio of the LIV to standard oscillation probabilities as a function of $L/E$ and $\dcp$. The reference energy $E=2.0$ GeV has been fixed to the \nova flux peak energy. For this peak energy and \nova baseline, $L/E = 405$ km/GeV. The upper (lower) panel shows the ratio for NH (IH), the left (right) panel shows it for neutrino (anti-neutrino). $|a_{e\mu}|$ and $|a_{e\tau}|$ have been fixed at $2\times 10^{-23}$ GeV each and we have set $\phi_{e\mu}=\phi_{e\tau}=0$.}}
\label{rat-2-0}
\end{figure}

\begin{figure}[htbp]
\centering
\includegraphics[width=1.0\textwidth]{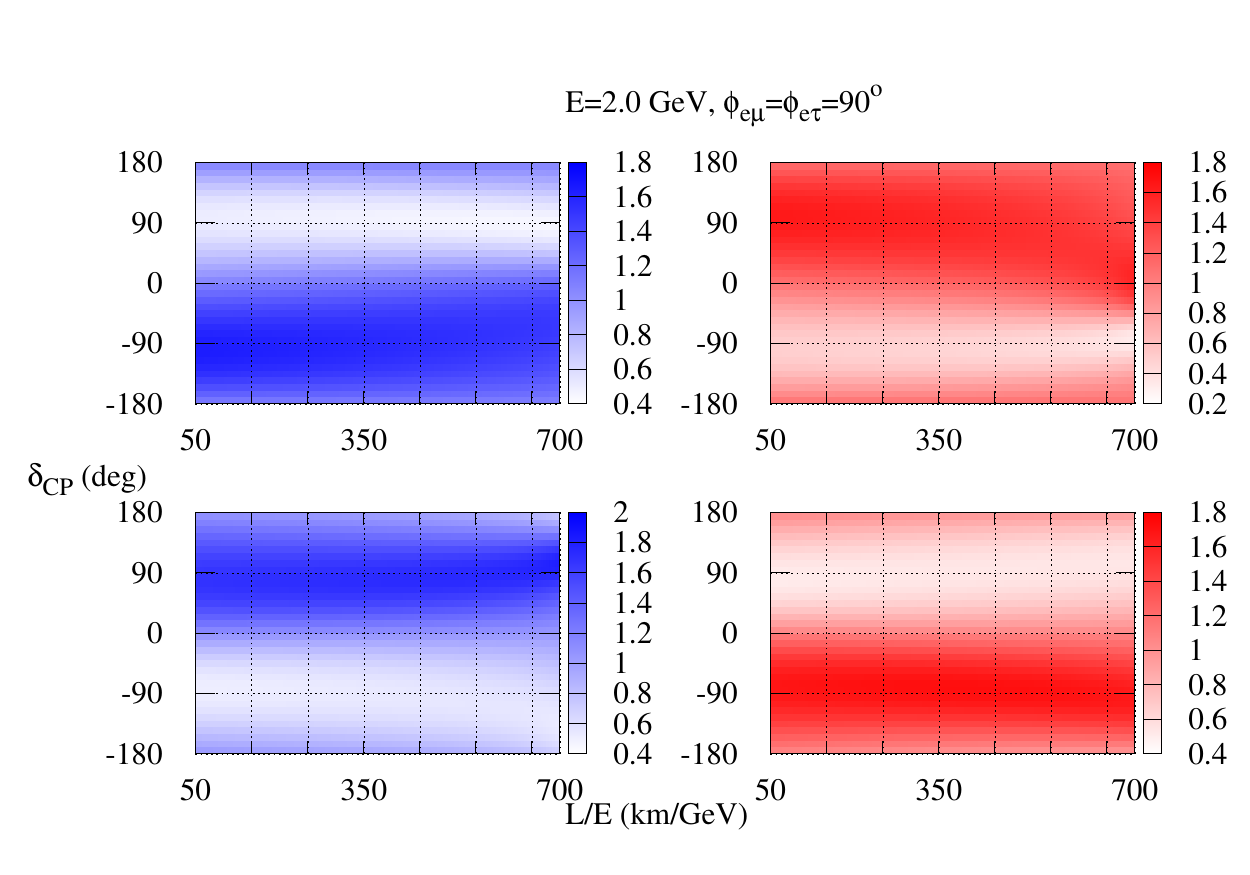}
\caption{\footnotesize{Ratio of the LIV to standard oscillation probabilities as a function of $L/E$ and $\dcp$. The reference energy $E=2.0$ GeV has been fixed to the \nova flux peak energy. For this peak energy and \nova baseline, $L/E = 405$ km/GeV. The upper (lower) panel shows the ratio for NH (IH), the left (right) panel shows it for neutrino (anti-neutrino). $|a_{e\mu}|$ and $|a_{e\tau}|$ have been fixed at $2\times 10^{-23}$ GeV each and we have set $\phi_{e\mu}=\phi_{e\tau}=90^\circ$.}}
\label{rat-2-90}
\end{figure}

\begin{figure}[htbp]
\centering
\includegraphics[width=1.0\textwidth]{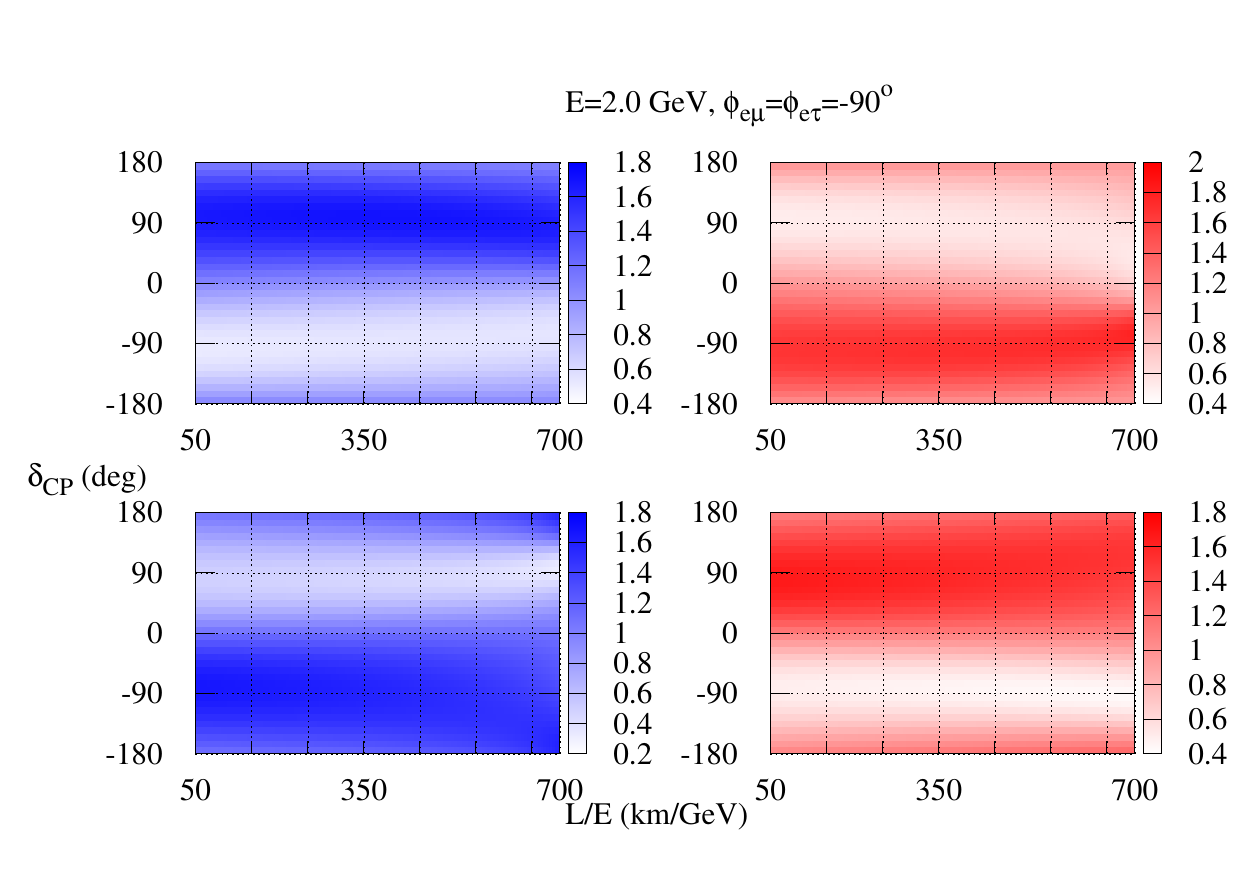}
\caption{\footnotesize{Ratio of the LIV to standard oscillation probabilities as a function of $L/E$ and $\dcp$. The reference energy $E=2.0$ GeV has been fixed to the \nova flux peak energy. For this peak energy and \nova baseline, $L/E = 405$ km/GeV. The upper (lower) panel shows the ratio for NH (IH), the left (right) panel shows it for neutrino (anti-neutrino). $|a_{e\mu}|$ and $|a_{e\tau}|$ have been fixed at $2\times 10^{-23}$ GeV each and we have set $\phi_{e\mu}=\phi_{e\tau}=-90^\circ$.}}
\label{rat-2--90}
\end{figure}

It would be interesting to note down the the ability to discriminate between the two models for different $\dcp$ values and baselines. To do so, we first fixed the neutrino energy at the \nova flux peak enegy $2.0$ GeV and varied $\dcp$ in the range $[-180^\circ:180^\circ]$ and the baseline from $100$ km to $1400$ km. In figures \ref{rat-2-0}-\ref{rat-2--90}, the ratio between $\pmue^{\rm SM+LIV}$ and $\pmue^{\rm SM}$ has been shown as a function of $L/E$ and $\dcp$ for different $\phi_{e\mu}$ and $\phi_{e\tau}$. values mentioned in the figure panels. $|a_{e\mu}|$ and $|a_{e\tau}|$ have been fixed at $2\times 10^{-23}$ GeV each. The standard oscillation parameters have been fixed at the current global best-fit values taken from \cite{nufit, Esteban:2020cvm}. The farther away the ratio is from 1, the better is the discrimination capability between the two models. It can be said that with $810$ km baseline and flux peak energy at $2.0$ GeV, \nova has a good discrimination capability between the two models at the oscillation probability level for all three cases and with both the hierarchies.

\begin{figure}[htbp]
\centering
\includegraphics[width=1.0\textwidth]{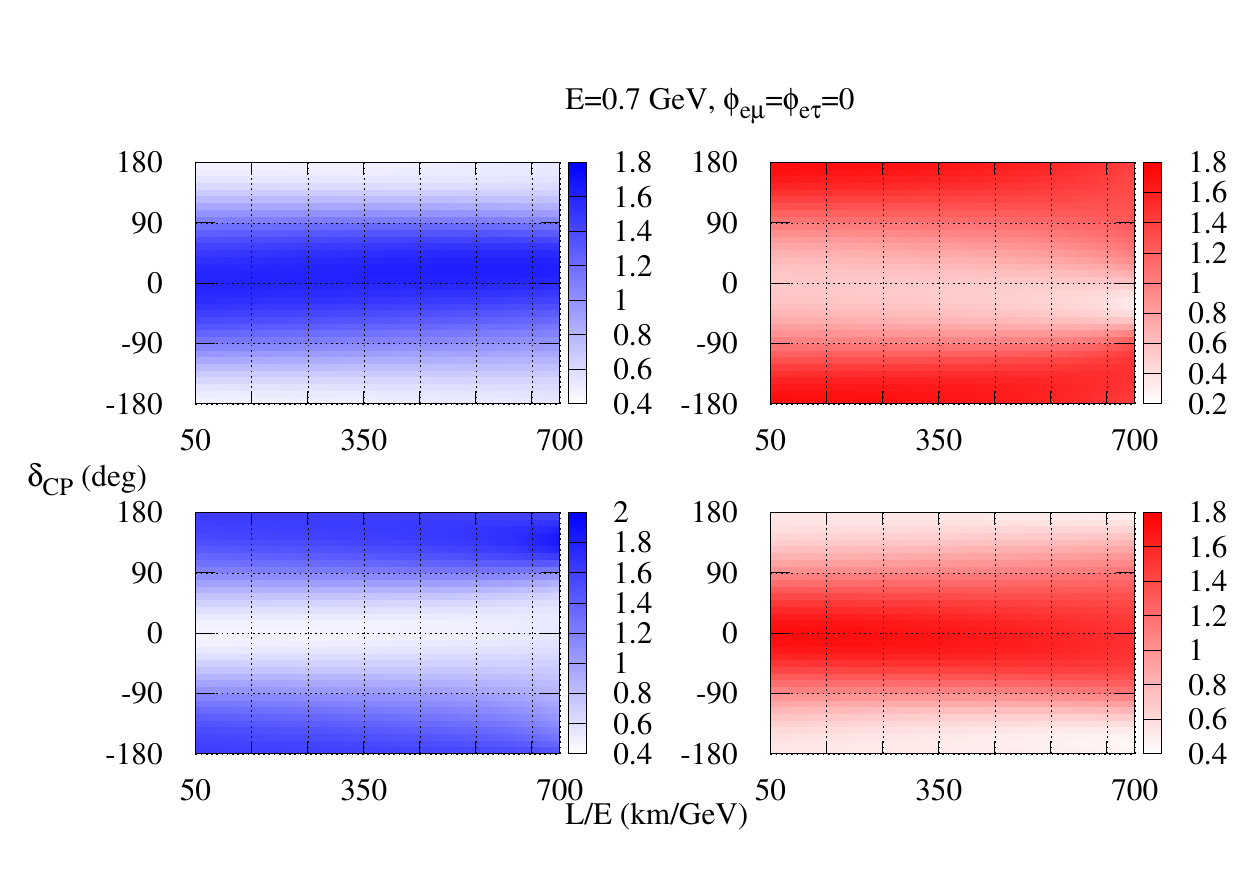}
\caption{\footnotesize{Ratio of the LIV to standard oscillation probabilities as a function of $L/E$ and $\dcp$. The reference energy $E=0.7$ GeV has been fixed to the T2K flux peak energy. For this peak energy and T2K baseline, $L/E = 421$ km/GeV. The upper (lower) panel shows the ratio for NH (IH), the left (right) panel shows it for neutrino (anti-neutrino). $|a_{e\mu}|$ and $|a_{e\tau}|$ have been fixed at $2\times 10^{-23}$ GeV each and we have set $\phi_{e\mu}=\phi_{e\tau}=0$.}}
\label{rat-0.7-0}
\end{figure}

\begin{figure}[htbp]
\centering
\includegraphics[width=1.0\textwidth]{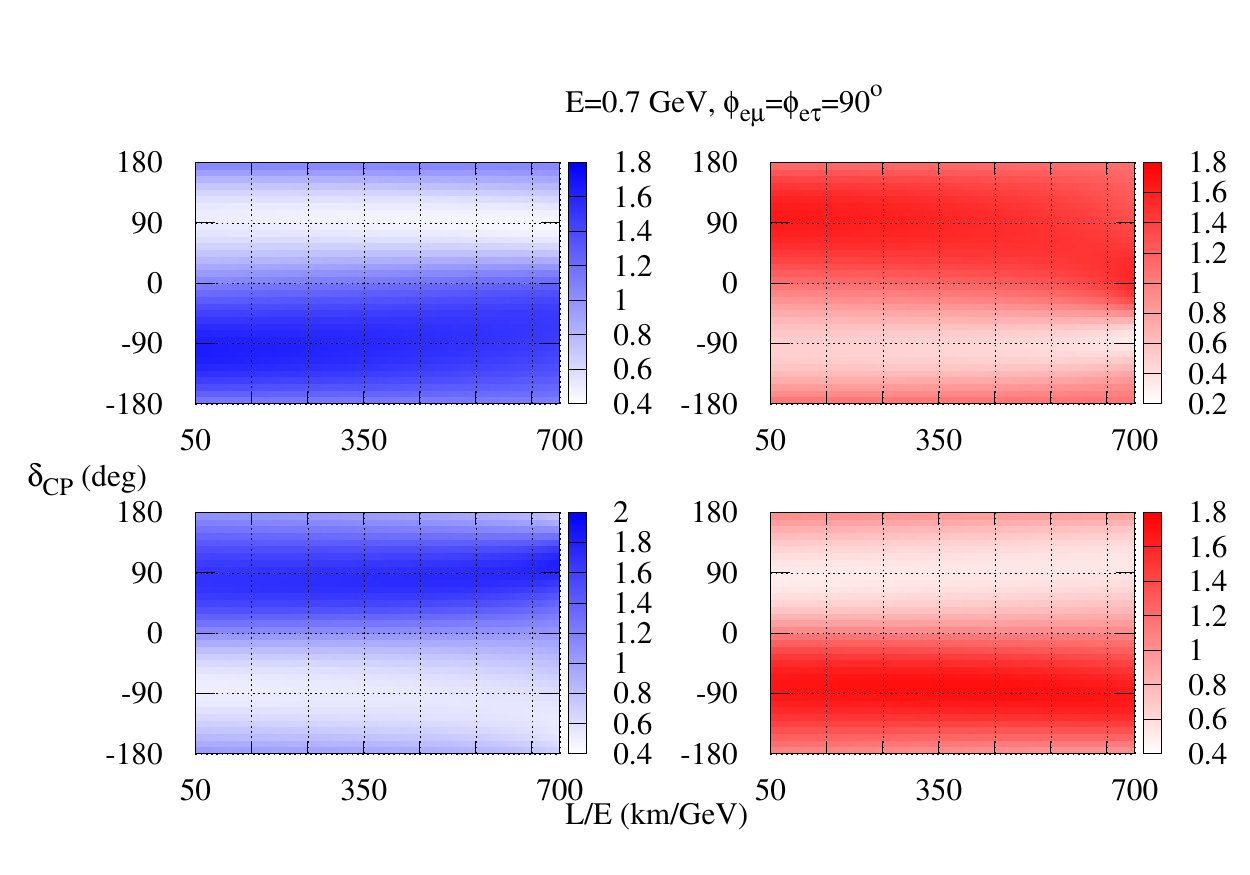}
\caption{\footnotesize{Ratio of the LIV to standard oscillation probabilities as a function of $L/E$ and $\dcp$. The reference energy $E=0.7$ GeV has been fixed to the T2K flux peak energy. For this peak energy and T2K baseline, $L/E = 421$ km/GeV. The upper (lower) panel shows the ratio for NH (IH), the left (right) panel shows it for neutrino (anti-neutrino). $|a_{e\mu}|$ and $|a_{e\tau}|$ have been fixed at $2\times 10^{-23}$ GeV each and we have set $\phi_{e\mu}=\phi_{e\tau}=90^\circ$.}}
\label{rat-0.7-90}
\end{figure}

\begin{figure}[htbp]
\centering
\includegraphics[width=1.0\textwidth]{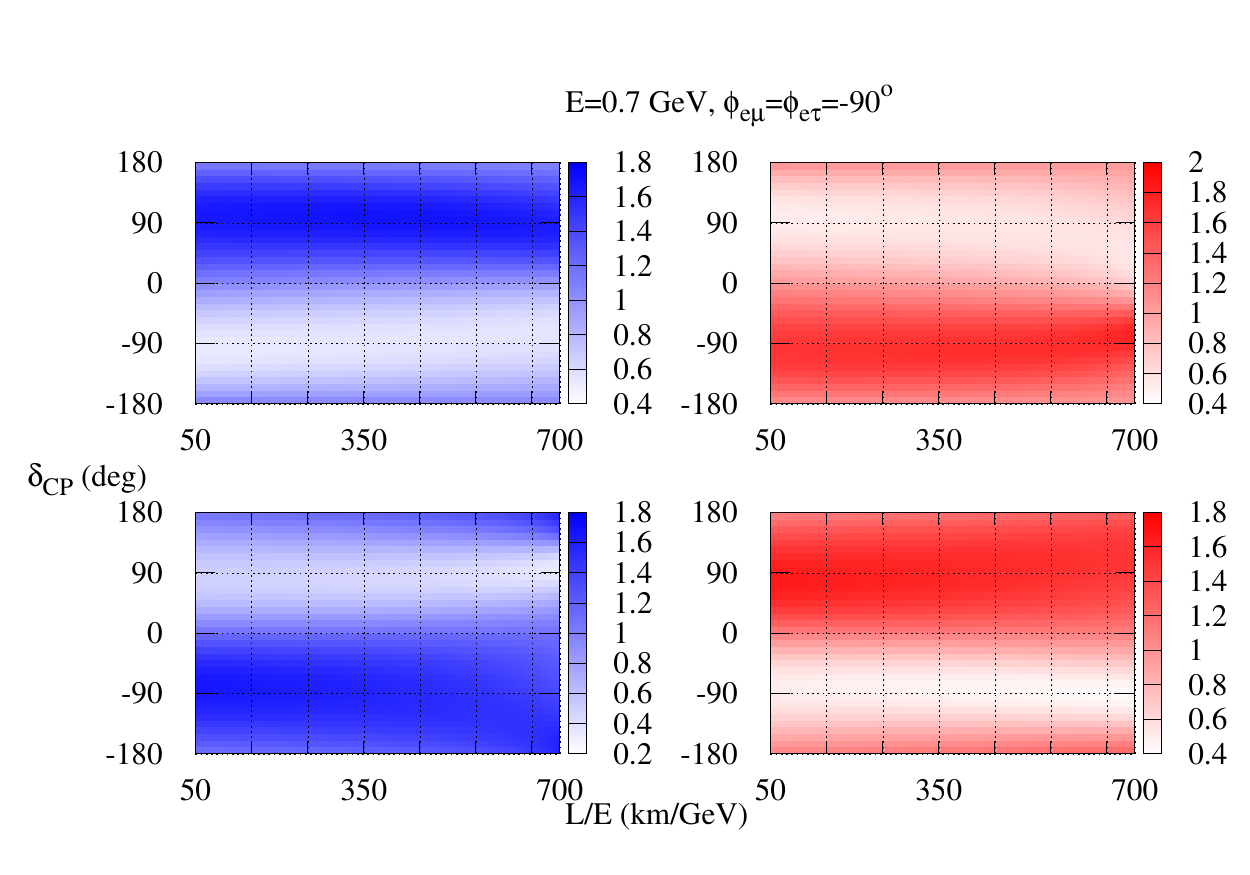}
\caption{\footnotesize{Ratio of the LIV to standard oscillation probabilities as a function of $L/E$ and $\dcp$. The reference energy $E=0.7$ GeV has been fixed to the T2K flux peak energy. For this peak energy and T2K baseline, $L/E = 421$ km/GeV. The upper (lower) panel shows the ratio for NH (IH), the left (right) panel shows it for neutrino (anti-neutrino). $|a_{e\mu}|$ and $|a_{e\tau}|$ have been fixed at $2\times 10^{-23}$ GeV each and we have set $\phi_{e\mu}=\phi_{e\tau}=-90^\circ$.}}
\label{rat-0.7--90}
\end{figure}

In the next step, we repeated the same exercise by fixing the energy $E=0.7$ GeV at the T2K flux peak energy. The results have been shown in figures \ref{rat-0.7-0}-\ref{rat-0.7--90}. It can be observed that T2K with its baseline of $L=295$ km, has a good discrimination capability at the probability level between the two models for most of the CP violating $\dcp$ values except for $\nu$ ($\bar{\nu}$) run when the hierarchy is IH (NH) and $\phi_{e\mu}=\phi_{e\tau}=0$.

\section{Analysis details}
\label{analysis}
The \nova detector \cite{Ayres:2004js} is a 14 kt totally active scintillator detector (TASD), placed 810 km away from the neutrino source at the Fermilab and it is situated at $0.8^\circ$ off-axis of the NuMI beam. The flux peaks at $2$ GeV, close to the oscillation maxima at 1.4 GeV for NH and at 1.8 GeV for IH. \nova started taking data in 2014 and took data that until 2020 release \cite{Himmel:2020}, correspond to $1.36 \times 10^{21}$ ($1.25 \times 10^{21}$) POT in $\nu$ ($\bar{\nu}$) mode.
The T2K experiment \cite{Itow:2001ee} uses the $\nu_\mu$ beam from the J-PARC accelerator at Tokai and the water Cherenkov detector at Super-Kamiokande, which is 295 km away from the source. The detector is situated $2.5^\circ$ off-axis. The flux peaks at $0.7$ GeV, which is also close to the first oscillation maximum.  T2K started taking data in 2009 and until 2020 release of results these \cite{Dunne:2020} correspond to $1.97 \times 10^{21}$ ($1.63 \times 10^{21}$) POT in $\nu$ ($\bar{\nu}$) mode.

To analyse the data, we kept $\sin^2\tx$ and $\ds$ to their best-fit values $0.304$ and $7.42\times 10^{-5}\, {\rm eV}^2$. These parameters have been determined by measuring electron neutrino survivavl probabilities in solar neutrino experiments. The LIV parameters which mostly affect electron neutrino survival probabilities are $a_{ee}$. We have fixed this parameter to zero. Therefore, the new physics should not affect the results from the solar neutrino experiments. We varied $\sin^2 2\ty$ in its $3\, \sigma$ range around its central value $0.084$ with $3.5\%$ uncertainty. $\sin^2 \tz$ has been varied in its $3\, \sigma$ range $[0.41:0.62]$. These ranges have been taken from the global best-fit ranges given in ref. \cite{Esteban:2018azc}. These global fit was done without the 2020 data from \nova and T2K. We varied $|\Delta_{\mu \mu}|$ in its $3\, \sigma$ range around the MINOS best-fit value $2.32\times 10^{-3}\, {\rm eV}^2$ with $3\%$ uncertainty \cite{Nichol:2012}. $\Delta_{\mu \mu}$ is related with $\dl$ by the following relation \cite{Nunokawa:2005nx}
\begin{equation}
\Delta_{\mu \mu}= \sin^2 \tz \dl + \cos^2 \tx \Delta_{32}+\cos \dcp \sin 2\tx \sin \ty \tan \tx \ds.
\end{equation}
The CP violating phase $\dcp$ has been varied in its complete range $[-180^\circ:180^\circ]$.

Among the new physics parameters, $|a_{e\mu}|$ and $|a_{e\tau}|$ have been varied in the range $0-20\times 10^{-23}$ GeV range. The phases $\phi_{e\mu}$ and $\phi_{e\tau}$ have been varied in the range $[-180^\circ:180^\circ]$. 

We calculated the theoretical event rates and the $\chi^2$ between data and theoretical event rates using GLoBES \cite{Huber:2004ka, Huber:2007ji}. The data has been taken from \cite{Himmel:2020, Dunne:2020}. To calculate the theoretical event rates, we fixed the signal and background efficiencies by matching with the Monte-Carlo simulations given by the collaborations \cite{Himmel:2020, Dunne:2020}. Automatic bin based energy smearing for generated theoretical events has been implemented in the same way as described in the GLoBES manual \cite{Huber:2004ka, Huber:2007ji}.
For this purpose, we used a Gaussian smearing function
\begin{equation}
R^c (E,E^\prime)=\frac{1}{\sqrt{2\pi}}e^{-\frac{(E-E^\prime)^2}{2\sigma^2(E)}},
\end{equation}
where $E^\prime$ is the reconstructed energy. The energy resolution function is given by 
\begin{equation}
\sigma(E)=\alpha E+\beta \sqrt{E}+\gamma.
\end{equation}
For NO$\nu$A, we used $\alpha=0.11\, (0.09)$, $\beta=\gamma=0$ for electron (muon) like events. For T2K, we used $\alpha=0$, $\beta=0.075$, $\gamma=0.05$ for both electron and muon like events. 

For NO$\nu$A, we used
\begin{itemize}
    \item $5\%$ normalisation and $5\%$ energy callibration systematics uncertainty for $e$ like events, and
    \item $5\%$ normalisation and $0.01\%$ energy callibration systematics uncertainty for $\mu$ like events. 
    
\end{itemize}
For T2K, we used 
\begin{itemize}
    \item $5\%$ normalisation and $5\%$ energy callibration systematics uncertainty for $e$ like events, and
    \item $5\%$ normalisation and $0.01\%$ energy callibration systematics uncertainty for $\mu$ like events.
\end{itemize}
Implementing systematics uncertainty has been discussed in details in GLoBES manual \cite{Huber:2004ka, Huber:2007ji}.
During the calculations of $\chi^2$ we added priors on $\sin^2 2\ty$. After calculating $\chi^2$, we found out minimum of these $\chi^2$s and subtracted it from all the $\chi^2$s to calculate $\dchsq$.
\section{Results and discussions}
\label{result}
\begin{figure}[htbp]
\centering
\includegraphics[width=1.0\textwidth]{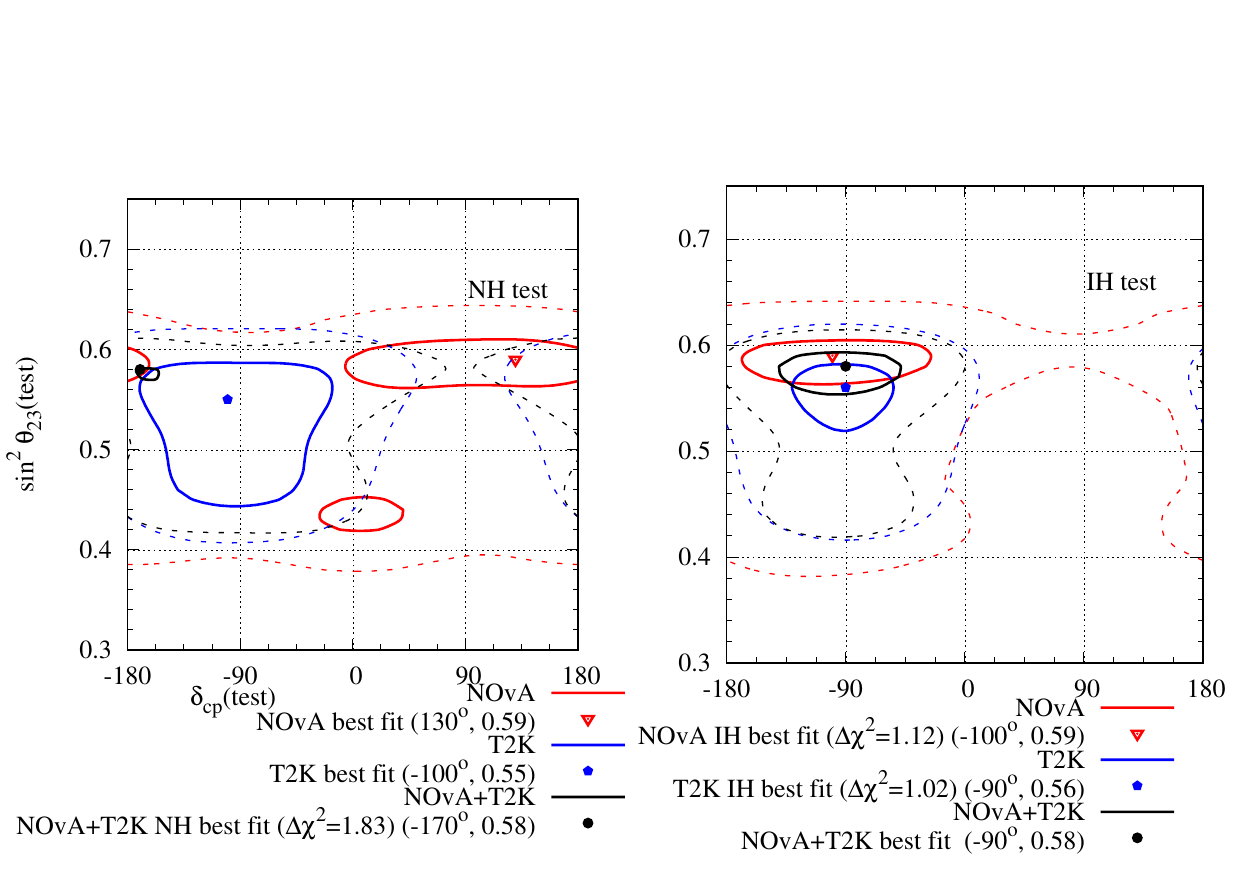}
\caption{\footnotesize{Allowed region in the $\sin^2 \tz-\dcp$ plane after analysing \nova and T2K complete data set with standard matter effect without LIV hypothesis. The left (right) panel represents test hierarchy NH (IH). The red (blue) lines indicate the results for \nova (T2K)
and the black line indicates the combined analysis of both. The solid (dashed) lines indicate the $1\, \sigma$ ($3\, \sigma$)
allowed regions. The minimum $\chi^2$ for \nova
(T2K) with 50 (88) bins is 48.65 (95.85) and it occurs at NH. For the combined analysis, the minimum $\chi^2$ with 138 bins is 147.14.}}
\label{standard}
\end{figure}

At first, we have analysed the data with the standard matter effect without any LIV hypothesis. The minimum $\chi^2$ for \nova (T2K) with 50 (88) bins is 48.65 (95.85) and it is at NH. For the combined analysis, the minimum $\chi^2$ with 138 bins is 147.14 and it occurs at IH. In fig.~\ref{standard}, we have shown the analysis in the $\sin^2 \tz-\dcp$ plane.
This plot is comparable to the ones presented by the collaborations in references \cite{Himmel:2020, Dunne:2020}. It is evident that there is a tension between the two experiments in terms of the best-fit $\dcp$ values. Moreover, there is no overlap between the $1\,\sigma$ region of each experiment.  

Once the standard analysis is done, we proceeded to analyse the data with LIV hypothesis. We found out that the minimum $\chi^2$ for \nova (T2K) with 50 (88) bins is 47.71 (93.14) and it is at NH. Both the experiments, however, have a degenerate solution at IH with $\dchsq=0.1$ For the combined analysis the minimum $\chi^2$ is 145.09 at IH for 138 bins. The combined analysis has a degenerate solution at NH with $\dchsq=0.1$. Therefore, the present accelerator neutrino oscillation data has no hierarchy sensitivity when analysed with LIV. 

\begin{figure}[htbp]
\centering
\includegraphics[width=1.0\textwidth]{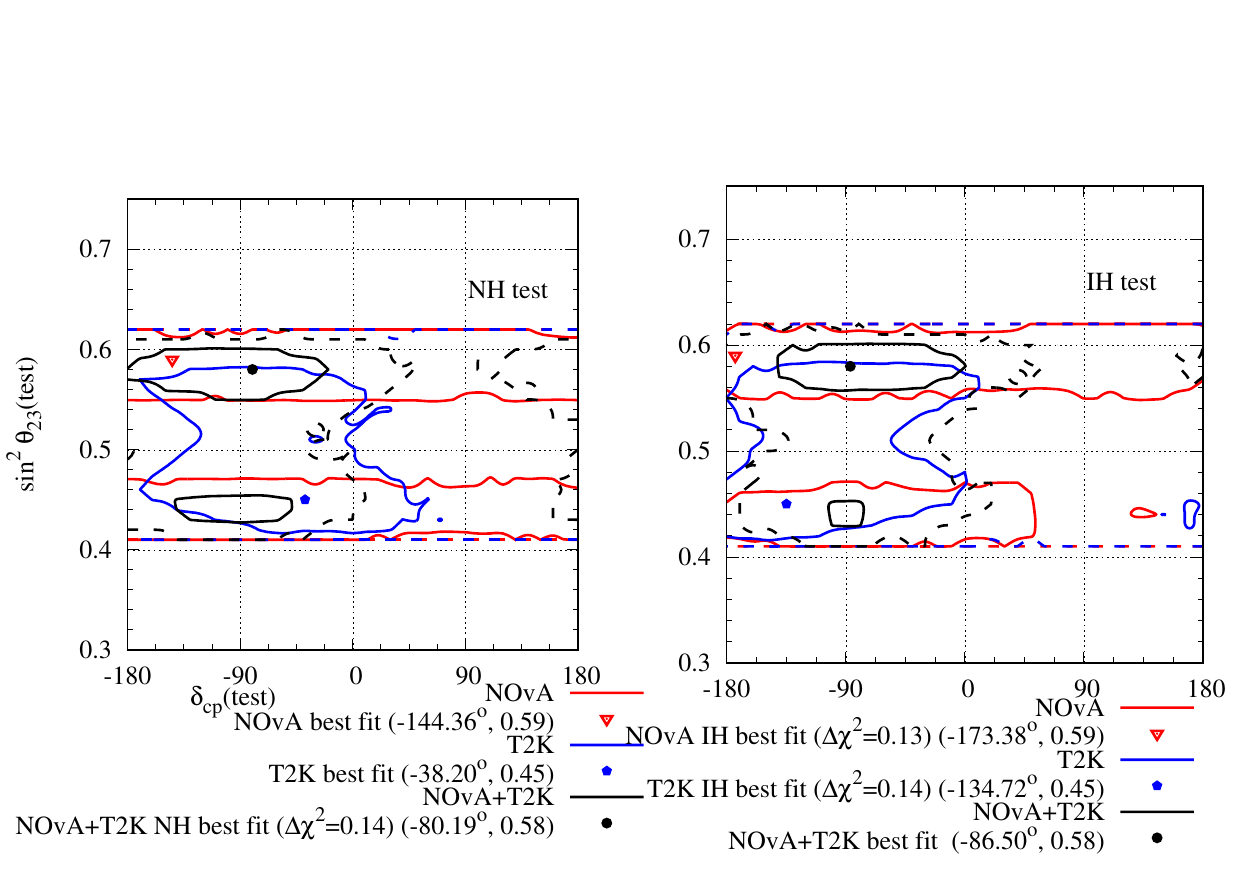}
\caption{\footnotesize{Allowed region in the $\sin^2 \tz-\dcp$ plane after analysing \nova and T2K complete data set with LIV hypothesis. The left (right) panel represents test hierarchy NH (IH). The red (blue) lines indicate the results for \nova (T2K)
and the black line indicates the combined analysis of both. The solid (dashed) lines indicate the $1\, \sigma$ ($3\, \sigma$)
allowed regions. The minimum $\chi^2$ for \nova
(T2K) with 50 (88) bins is 47.71 (93.14) and it occurs at NH. For the combined analysis, the minimum $\chi^2$ with 138 bins is 145.09.}}
\label{LIV}
\end{figure}

\begin{figure}[htbp]
\centering
\includegraphics[width=0.5\textwidth]{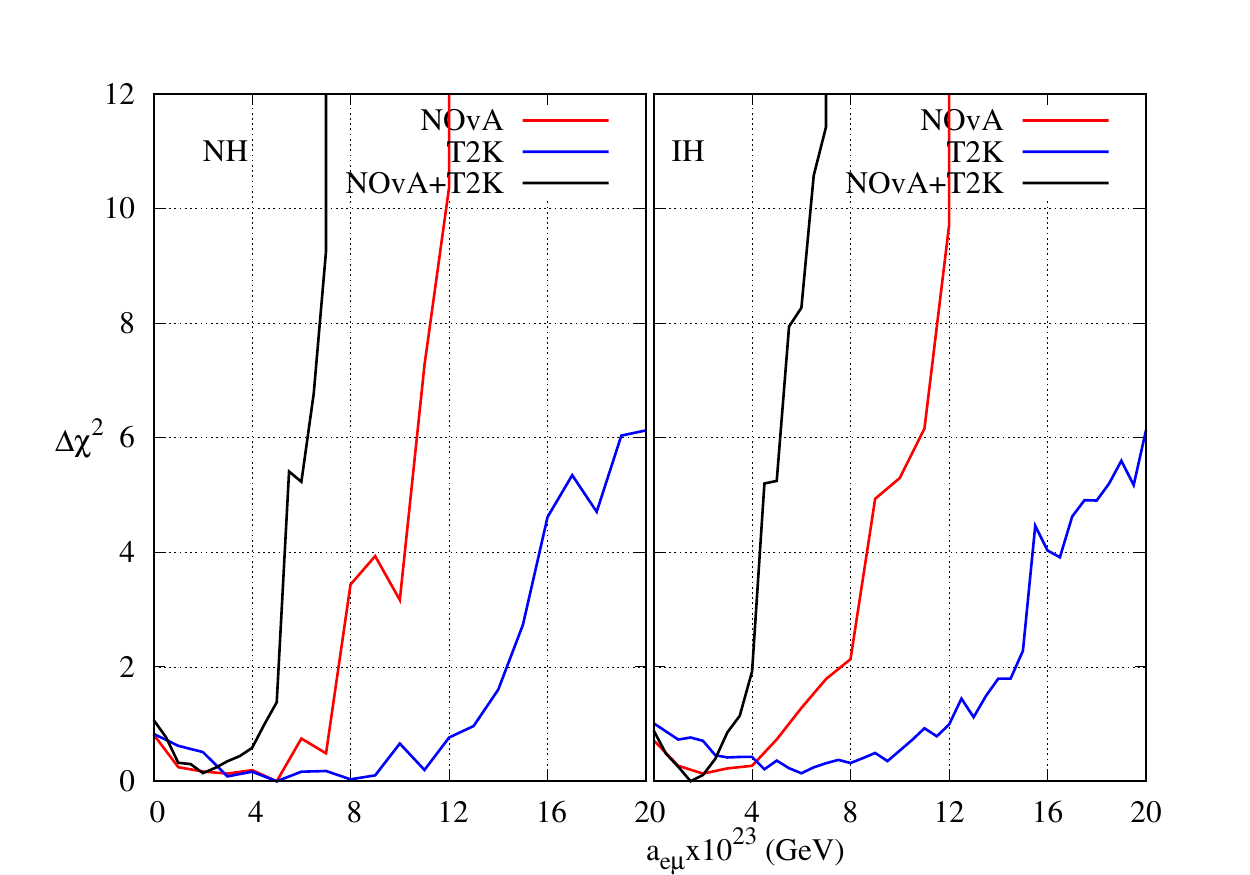}
\includegraphics[width=0.5\textwidth]{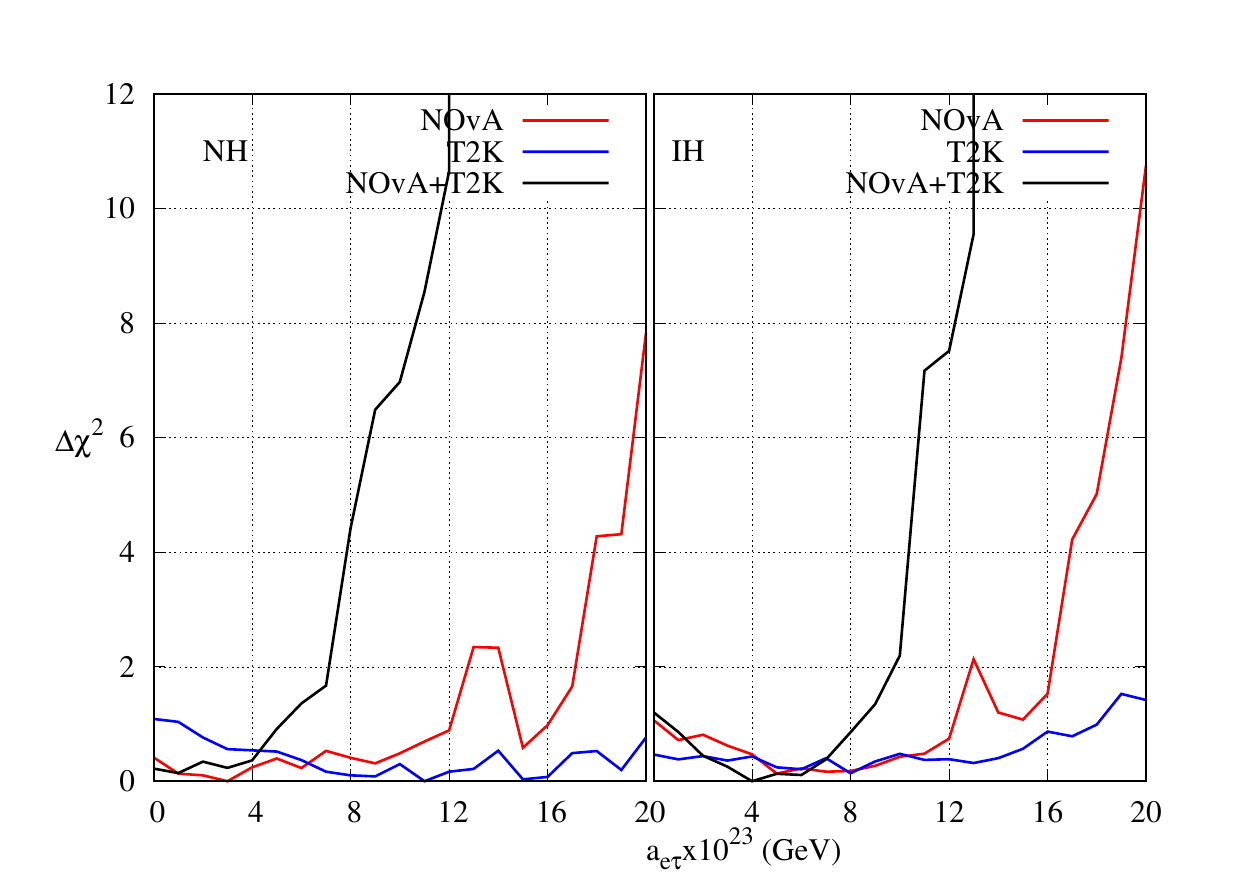}
\includegraphics[width=0.5\textwidth]{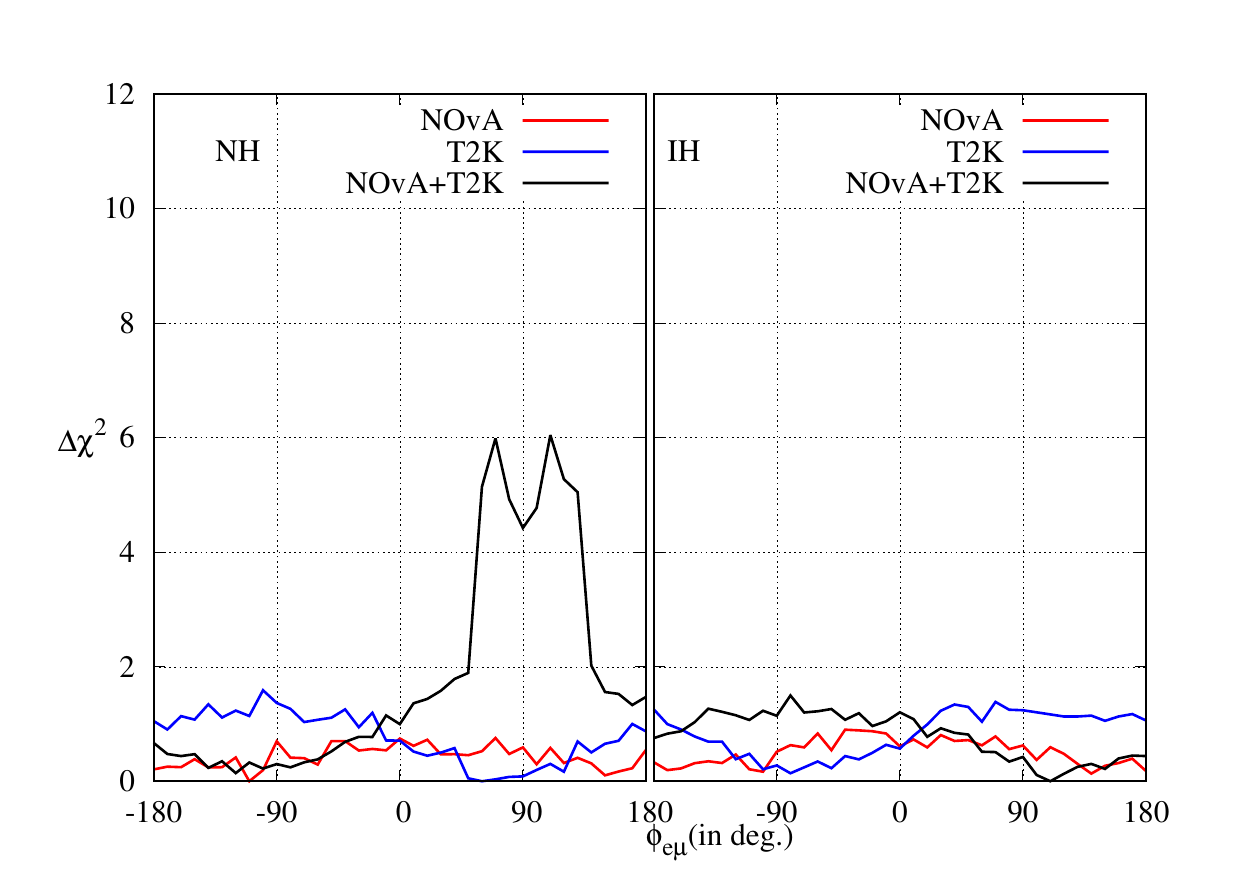}
\includegraphics[width=0.5\textwidth]{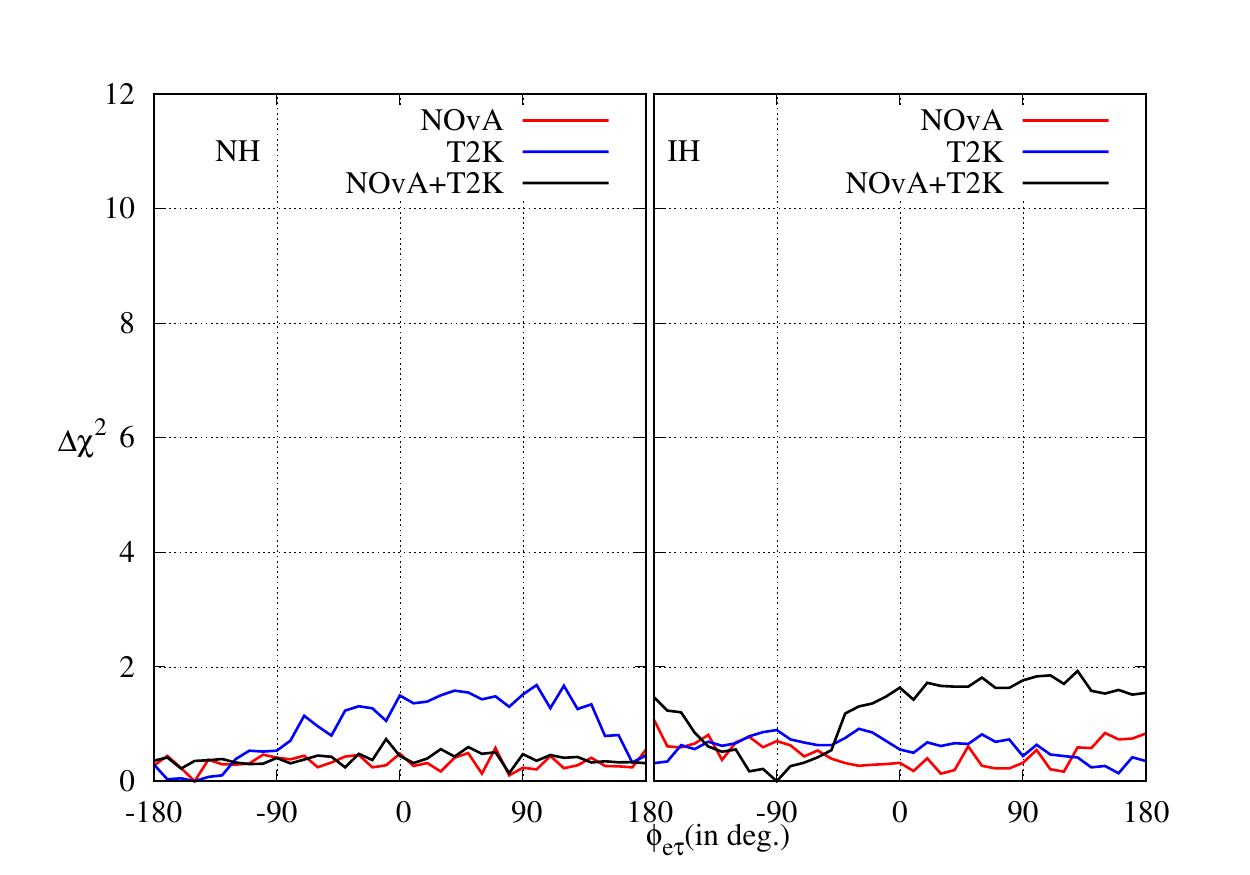}
\caption{\footnotesize{$\dchsq$ as a function of individual LIV parameters.}}
\label{param}
\end{figure}

In fig.~\ref{LIV}, we have presented our result with LIV hypothesis on the $\sin^2\tz-\dcp$ plane. T2K (NO$\nu$A) disfavours (includes) \nova (T2K) best-fit points at $1\, \sigma$ C.L. Now, there is a large overlap between the $1\, \sigma$ allowed regions of the two experiments. Hence, one can conclude that the tension between the two experiments gets reduced when the data are analysed with LIV. However, there is a new mild tension in terms of the best-fit $\sin^2\tz$ values. $\tz$ at T2K best-fit point is at lower octant, while the same is at higher octant for NO$\nu$A. But \nova has a nearly degenerate ($\dchsq=0.35$) best-fit point at lower octant. Similarly, T2K also cannot rule out higher octant at $1\, \sigma$ C.L. Thus, both the experiments lose their octant sensitivity when analysed with LIV. 
In fig.~\ref{param}, we have shown $\dchsq$ as a function of the individual LIV parameters. To do so, we marginalised $\dchsq$ on all the parameters except the one against which we want to plot it. It can be seen from fig.~\ref{param} that both $|a_{e\mu}|=0$ and $|a_{e\tau}|=0$ values have $\dchsq>1$ for T2K. For \nova however, both these values have $\dchsq<1$. For the combined analysis, $|a_{e\mu}|=0$ has a $\dchsq>1$, but $|a_{e\tau}|=0$ has $\dchsq<1$. Therefore, it can be said that the present \nova data do not favour any of the two hypotheses over the other. However T2K data and the combined analysis disfavour standard oscillation at $1\, \sigma$ C.L. In tables \ref{nova}, \ref{t2k} and \ref{nova+t2k}, we have presented the best-fit values of the standard and non-standard unknown oscillation parameters.

\begin{figure}[htbp]
\centering
\includegraphics[width=1.0\textwidth]{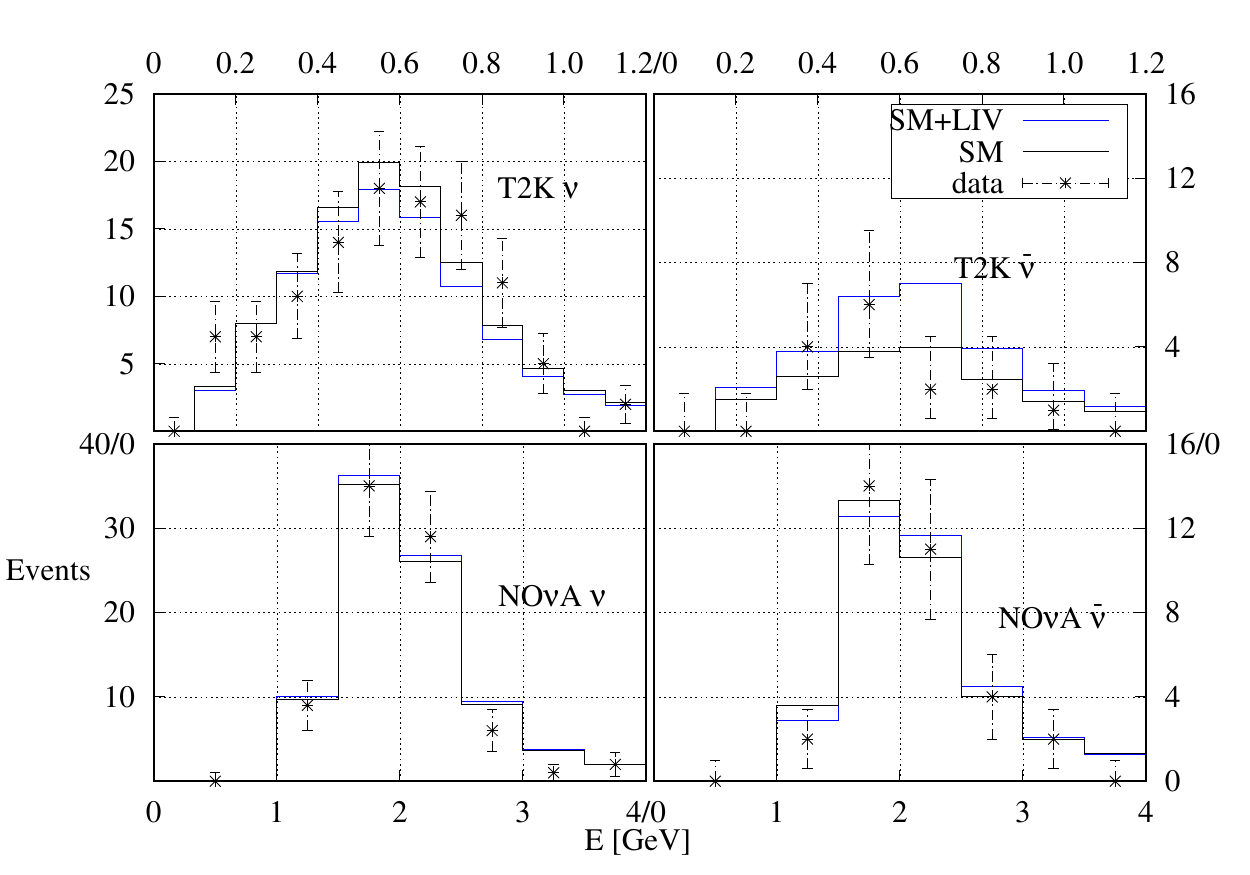}
\caption{\footnotesize{Expected binned electron and positron event rates as a function of energy at the best-fit points for both the models and for both \nova and T2K experiments. The experimental event rates have also been plotted.}}
\label{events}
\end{figure}

To emphasize our result, in fig.~\ref{events}, we have presented the expected electron and positron events rates for each energy bins for both standard oscillation and oscillation with LIV as a function of energy for both \nova and T2K. The experimental event rates have also been plotted. It is obvious that for NO$\nu$A, there is not any significant difference between the expected event rates at best-fit points of the two models and both models give a good fit to the data. However, for T2K, there is a clear distinction at the expected event rates at the best-fit points of the two models. Also, LIV gives a better fit to the data especially at energies close to the flux peak energy.

\begin{table}

%\hspace* {-15 mm}
%\tbl{Parameter values at the best-fit points for NO$\nu$A.The $1\, \sigma$ error bars have been mentioned where possible. The $90\%$ limits for 1 d.o.f have also been mentioned. In each box, the result with 2019 (2020) data has been mentioned at the top (bottom) of the box. }
{\footnotesize
  \begin{tabular}{|l|l|l|l|l|l|l|}
    \hline
    {Parameters} &
      \multicolumn{2}{c|}{standard} &
      \multicolumn{2}{c|}{LIV} &
       \multicolumn{2}{c|}{$90\%$} \\

    & NH & IH  & NH & IH & NH & IH \\
    \hline
     $\frac{\Delta_{\mu \mu}}{10^{-3}\, {\rm eV}^2}$  & $2.44^{+0.02}_{-0.048}$ & $-(2.44^{+0.02}_{-0.048})$ & $2.40_{-0.026}^{+0.004}$& $-(2.41_{-0.05}^{+0.01})$ && \\
     
     \hline
     $\sin^2 2\ty$ & $0.084_{-0.002}^{+0.002}$ & $0.084_{-0.002}^{+0.003}$ &  $0.084_{-0.003}^{+0.002}$ & $0.084_{-0.003}^{+0.002}$ && \\
     \hline
     
   $\sin^2 \tz$
   &$0.59^{+0.01}_{-0.01}$&$0.59^{+0.01}_{-0.02}$&$0.43^{+0.03}_{-0.02}\oplus 0.59^{+0.03}_{-0.03}$&$0.43^{+0.02}_{-0.01}\oplus0.59^{+0.03}_{-0.03}$&&\\
   
    \hline

    $\dcp/^\circ$ 
    &$130^{+40}_{-110}$&$-(100^{+50}_{-60})$&$-144.36$&$-173.38$&&\\
    
    \hline
   
    $\frac{|a_{e\mu}|}{10^{-23}{\rm GeV}}$
    &&&$4.81$&$2.22$&$<8.19$& $<7.78$\\
    \hline
    $\frac{|a_{e\tau}|}{10^{-23}{\rm GeV}}$
    &&&$2.52$&$5.33^{+9.20}_{-5.33}$&$<3.18$&$<15.71$ \\
    
    \hline
    $\phi_{e\mu}$ & & & $-114.52$ & $141.18$ &   & \\

    \hline
    $\phi_{e\tau}/^\circ$ 
    
    &&&$-145.02$&$25.04$&&\\
    
    \hline
  \end{tabular}
  }
  \caption{Parameter values at the best-fit points for NO$\nu$A.The $1\, \sigma$ error bars have been mentioned where possible. The $90\%$ limits for 1 d.o.f have also been mentioned.}
  \label{nova}
\end{table}

 \begin{table}
%\hspace* {-15 mm}
%\tbl{Parameter values at the best-fit points for T2K.The $1\, \sigma$ error bars have been mentioned where possible. The $90\%$ limits for 1 d.o.f have also been mentioned. In each box, the result with 2019 (2020) data has been mentioned at the top (bottom) of the box.}
{\footnotesize
  \begin{tabular}{|l|l|l|l|l|l|l|}
    \hline
    {Parameters} &
      \multicolumn{2}{c|}{standard} &
      \multicolumn{2}{c|}{LIV} &
       \multicolumn{2}{c|}{$90\%$} \\
      % \multicolumn{2}{c|}{non-unitary ($>3\, \sigma$)} \\

    & NH& IH  & NH & IH  & NH &IH \\
    
    \hline
     
      $\frac{\Delta_{\mu \mu}}{10^{-3}\, {\rm eV}^2}$  &  $2.512^{+0.048}_{-0.048}$ & $-(2.512_{+0.048}^{-0.048})$ & $2.462_{-0.04}^{+0.04}$& $-(2.47_{-0.03}^{+0.05})$ & &\\
     
     \hline
     $\sin^2 2\ty$ & $0.084_{-0.002}^{+0.002}$ & $0.084_{-0.002}^{+0.003}$ &  $0.084_{-0.003}^{+0.002}$ & $0.084_{-0.003}^{+0.002}$ && \\
     \hline
     
   $\sin^2 \tz$&$0.55^{+0.03}_{-0.09}$&$0.56^{+0.02}_{-0.03}$&$0.45^{+0.03}_{-0.03}$&$0.45^{+0.03}_{-0.03}$&&\\
   
    \hline

    $\dcp/^\circ$&$-(100^{+50}_{-60})$&$-(90^{+30}_{-30})$&$-(38.20^{+65.71}_{-66.04})$&$-(134.72^{+45.28}_{-70.57})$&&\\
    
    \hline
   
    $\frac{|a_{e\mu}|}{10^{-23}{\rm GeV}}$
    &&&$4.60$&$6.17_{-6.02}^{+4.86}$&$<15.25$& $<14.92$\\
    \hline
    $\frac{|a_{e\tau}|}{10^{-23}{\rm GeV}}$ 
    &&&$11.14$&$8.06$&Out of range&Out of range \\
    
    \hline
    $\phi_{e\mu}$ 
    &&&$64.29$&$-77.34$&&\\
    
    \hline
    $\phi_{e\tau}/^\circ$ 
    &&&$-153.24$&$158.07$&&\\
    
    \hline
  \end{tabular}
  }
  \caption{Parameter values at the best-fit points for T2K.The $1\, \sigma$ error bars have been mentioned where possible. The $90\%$ limits for 1 d.o.f have also been mentioned.}
  \label{t2k}
\end{table}

 \begin{table}
%\hspace* {-15 mm}
%\tbl{Parameter values at the best-fit points for the combined analysis of NO$\nu$A and T2K.The $1\, \sigma$ error bars have been mentioned where possible. The $90\%$ limits for 1 d.o.f have also been mentioned. In each box, the result with 2019 (2020) data has been mentioned at the top (bottom) of the box.}
{\footnotesize
  \begin{tabular}{|l|l|l|l|l|l|l|}
    \hline
    {Parameters} &
      \multicolumn{2}{c|}{standard} &
      \multicolumn{2}{c|}{LIV} &
       \multicolumn{2}{c|}{$90\%$} \\
      % \multicolumn{2}{c|}{non-unitary ($>3\, \sigma$)} \\

    & NH& IH  & NH & IH  & NH &IH \\
    \hline
      $\frac{\Delta_{\mu \mu}}{10^{-3}\, {\rm eV}^2}$  &  $2.464^{+0.024}_{-0.048}$ & $-(2.464^{+0.024}_{-0.048})$& $2.462_{-0.04}^{+0.02}$& $-(2.462_{-0.023}^{+0.003})$ & &\\
     
     \hline
     $\sin^2 2\ty$ & $0.084_{-0.002}^{+0.002}$ & $0.084_{-0.002}^{+0.003}$ &  $0.084_{-0.003}^{+0.002}$ & $0.084_{-0.003}^{+0.002}$ && \\
     \hline
   $\sin^2 \tz$& $0.55^{+0.03}_{-0.09}$&$0.56^{+0.02}_{-0.03}$&$0.44^{+0.02}_{-0.01}\oplus 0.58^{+0.02}_{-0.02}$&$0.58^{+0.02}_{-0.02}$&&\\
   
    \hline

    $\dcp/^\circ$&$-(100^{+50}_{-60})$&$-(90^{+30}_{-30})$&$-(80.19^{+69.80}_{-36.21})$&$-(86.50^{+38.62}_{-54.67})$&&\\
    
    \hline
   
    $\frac{|a_{e\mu}|}{10^{-23}{\rm GeV}}$ &&&$1.86^{+2.57}_{-1.86}$&$1.52$&$<4.80$& $<3.80$\\
    \hline
    $\frac{|a_{e\tau}|}{10^{-23}{\rm GeV}}$ & & & $0.57$ & $4.16$ & $<6.70$& $<9.50$\\

    \hline
    $\phi_{e\mu}/^\circ$ & & & $-115.72$ & $110.08$ & $$  & $$\\

    \hline
    $\phi_{e\tau}/^\circ$ & & & $84.36$ & $-89.27$ & & \\

    \hline
  \end{tabular}
  }
  \caption{Parameter values at the best-fit points for the combined analysis of NO$\nu$A and T2K.The $1\, \sigma$ error bars have been mentioned where possible. The $90\%$ limits for 1 d.o.f have also been mentioned.}
  \label{nova+t2k}
\end{table}

\section{Conclusion}
At present, \nova cannot disfavour any of the two hypotheses. However, both T2K and the combined analysis disfavours standard oscillation at $1\, \sigma$ C.L. The latest accelerator neutrino oscillation data lose hierarchy sensitivity when analysed with LIV. The $1\, \sigma$ allowed regions from the two experiments have a large overlap with LIV, unlike the standard oscillation case. Therefore, one can comment that the tension between the two experiments is reduced when the data are analysed with LIV. However, there is a new mild tension between the best-fit values of $\sin^2 \tz$. While $\tz$ for \nova best-fit is at HO, it is at LO for the T2K bst-fit point. But \nova (T2K) has a nearly degenerate best-fit point at lower (higher) octant as well. Therefore, both \nova and T2K data do not have octant and hierarchy determination capability when analysed with LIV.

In light of these, we recommend that the long baseline accelerator neutrino experiments to be analysed with LIV along with other BSM physics. If the future data continue to favour LIV over standard oscillation, it can be considered as a prominent signal for LIV. 
\label{conclusion}
 \section{Acknowledgement}
U.R. thanks Prof. Soebur Razzaque for the valuable comments.
\bibliographystyle{apsrev}
\bibliography{referenceslist}
\end{document}